\documentclass[%
reprint,
superscriptaddress,
 amsmath,amssymb,
 aps,
 pre
]{revtex4-2}

\usepackage{graphicx}
\usepackage{dcolumn}
\usepackage{bm}
\usepackage{physics}
\usepackage[colorlinks=true,linkcolor=blue,urlcolor=blue,citecolor=blue]{hyperref}

\begin{document}

\title{Hydrated porous substrates modulate the spatiotemporal dynamics of bacterial turbulence}

\author{Zixiang Lin}
\thanks{These two authors contributed equally.}
\affiliation{Global College, Shanghai Jiao Tong University, Shanghai 200240, People’s Republic of China}
\author{Jin Zhu}
\thanks{These two authors contributed equally.}
\affiliation{School of Physics, Georgia Institute of Technology, Atlanta, GA 30332, USA}
\affiliation{Interdisciplinary Graduate Program in Quantitative Biosciences, Georgia Institute of Technology, Atlanta, GA 30332, USA}
\author{Hongyi Bian}
\affiliation{Global College, Shanghai Jiao Tong University, Shanghai 200240, People’s Republic of China}
\author{Zijie Qu}
\email{zijie.qu@sjtu.edu.cn}
\affiliation{Global College, Shanghai Jiao Tong University, Shanghai 200240, People’s Republic of China} 


\begin{abstract}
Boundary conditions strongly influence bacterial turbulence, motivating the question of how hydrated porous substrates affect near-interface collective motion. We characterize the spatiotemporal dynamics of dilute and dense \textit{Escherichia coli} suspensions above agar substrates of varying concentrations, with glass serving as a nonporous reference. In dense suspensions shortly after loading, collective in-plane kinetic energy varies nonmonotonically with agar concentration, while individual swimming energy and characteristic spatial scales show comparatively little variation. Over five minutes above agar, the collective flow attenuates strongly, accompanied by shorter spatial correlations, longer temporal correlations, and a greater relative spectral weight at higher wavenumbers. To explore how the porous boundary may contribute to these changes, we extend mean-field kinetic theory (MFKT) to incorporate normal seepage and tangential slip at the interface. Linear stability analysis distinguishes near-interface collective modes from those in the bulk and identifies how fluid-layer thickness and permeability-dependent slip length modify the instability cutoff. The predicted reduction in the associated length scale provides a qualitative interpretation of the smaller flow structures observed over time and at higher agar concentrations. These findings reveal distinct substrate-dependent changes in the intensity, persistence, and spatial organization of near-interface bacterial turbulence. 
\end{abstract}

\maketitle

\textit{Introduction}.---Boundary conditions play a crucial role in modulating collective flows in active matter, as fluid motion driven by internally generated stresses remains hydrodynamically coupled to its surroundings, where momentum is transmitted and dissipated \cite{Marchetti2013, annurev-ActiveTurbulence}. 
Dense bacterial suspensions provide a striking example, developing spatiotemporal chaotic, vortex-rich flows at low Reynolds numbers \cite{Sokolov2007, Sokolov2012, Wensink2012, Dunkel2013, Dunkel2013M}, while keeping highly sensitive to boundaries \cite{Li2024}.
Geometric confinements, such as channels \cite{Wioland2016, Henshaw2023}, circular wells \cite{Wioland2013, Beppu2021, puggioni_giant_2022, nishiguchi_vortex_2025}, pillar lattices \cite{Nishiguchi2018, Reinken2020}, and parallel walls \cite{Wei2024}, can select coherent vortices and modify turbulent length scales. 
Beyond this, substrate friction and fluid interfaces alter the dissipation and transmission of active flows \cite{Boffetta2005, thijssen_anisotropic_2020, mecke_chiral_2024, MartnezPrat2021, spera_lowpass_2026}. 
However, how such flows respond to a hydrated porous substrate, which can permit both normal fluid exchange and tangential slip, remains poorly understood.

Hydrated agar provides a natural setting for addressing this question, as its concentration is known to influence bacterial collective motion on semisolid surfaces \cite{zhang_swarming_2009, kearns_field_2010, beer_statistical_2019}. 
In swarming colonies, however, collective motion is coupled to surfactant-mediated wetting, liquid redistribution, and cell–substrate interactions \cite{beer_colony_2009, pokhrel_friction_2025}, making it difficult to isolate the hydrodynamic response of the suspension itself. 
Moreover, bacterial swarming typically occurs in thin liquid layers, whereas bacterial turbulence can develop in thicker suspensions involving three-dimensional motion \cite{peng_imaging_2021, liu_density_2021}. 
Studying bacterial turbulence above agar therefore offers a way to examine the role of hydrated substrates in collective motion while recognizing that changes in agar concentration may affect several interfacial and fluid properties simultaneously.


\begin{figure*}[htb]
    \centering
    \includegraphics[width=\textwidth]{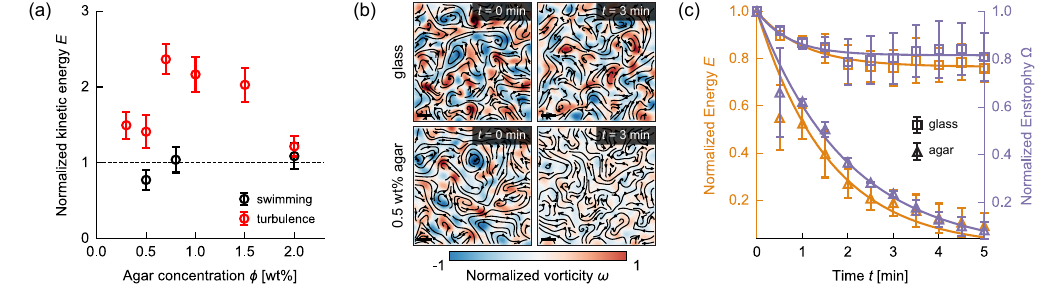}
    \caption{Hydrated porous substrates exert significant modulation on bacterial turbulence. (a) Kinetic energy $E$ versus agar concentration $\phi$. The kinetic energy $E$ is normalized by the mean value measured over the glass substrates. The reference kinetic energies are $E^{\text{Ref}}_{\text{swim}} = 362.3~\mathrm{\mu m^2/s^2}$ for the dilute suspension, and $E^{\text{Ref}}_{\text{turb}} = 78.7~\mathrm{\mu m^2/s^2}$ for the dense suspension. (b) Snapshots of flow fields over the glass (top) and 0.5~wt\% agar (bottom) at $t = 0$ (left) and $t = 3 \min$ (right). Streamlines are denoted by black lines with arrows, and colors indicate the vorticity normalized by its maximum magnitude at $t=0$.  Scale bars, 20~$\mathrm{\mu m}$. (c) Normalized in-plane kinetic energy $E$ and enstrophy $\Omega$ during the observation period. Squares denote glass and triangles denote agar. Solid lines represent exponential fits. Error bars indicate the standard error.}
    \label{fig:modulation}
\end{figure*}

Here, we investigate bacterial turbulence above agar substrates of varying concentrations, using glass as a nonporous reference. By comparing dilute and dense \textit{Escherichia coli} (\textit{E. coli}) suspensions, we distinguish substrate-dependent changes in individual swimming from those in collective motion. Shortly after loading, collective kinetic energy varies nonmonotonically with agar concentration, peaking near 0.7 wt\%, even though individual swimming speed and characteristic flow scales change comparatively little. Over the following five minutes, collective flows above agar attenuate strongly, in contrast to the comparatively persistent motion above glass. Remarkably, this attenuation is accompanied by shorter spatial correlations but longer temporal correlations, while normalized energy spectra acquire greater relative weight at higher wavenumbers. To interpret these observations, we extend an existing mean-field kinetic theory (MFKT) theoretical framework \cite{kultty2020, kultty2024, Brdfalvy2024}, to incorporate normal seepage and tangential slip at the porous interface. Linear stability analysis predicts that decreasing the fluid-layer thickness or slip length shifts the interfacial instability cutoff toward higher wavenumbers, consistent in trend with the observed reduction in spatial correlation length. Together, these results show that hydrated porous substrates modify the strength and spatiotemporal organization of bacterial turbulence in distinct ways, while suggesting interfacial hydrodynamic coupling as a mechanism contributing to its spatial response.


\begin{figure*}[tb]
    \centering
    \includegraphics[width=\textwidth]{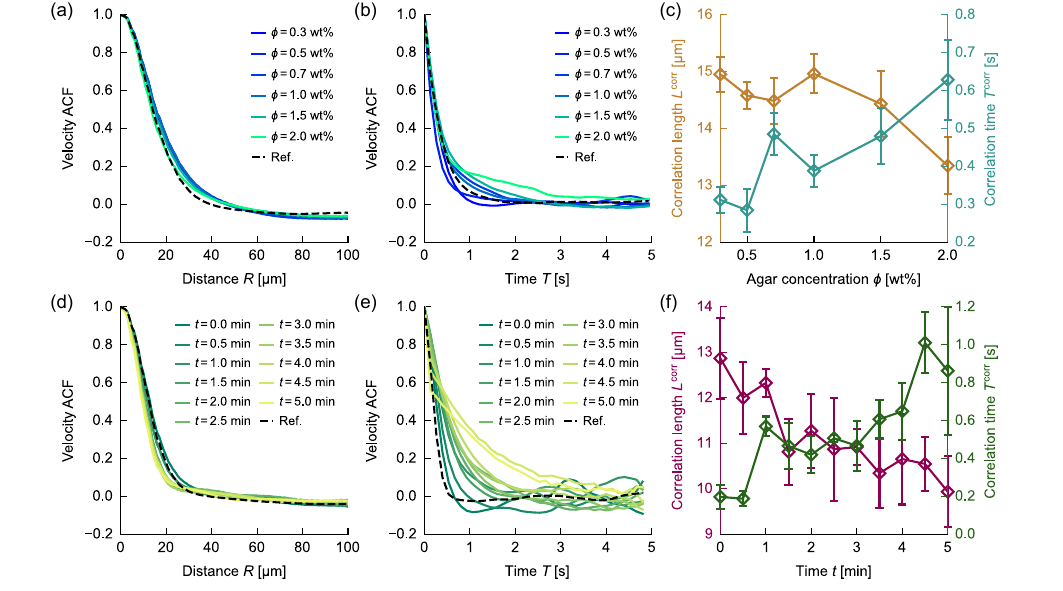}
    \caption{Variation in spatiotemporal structure of bacterial turbulence with agar concentration and time. (a) Spatial velocity ACF slightly narrows as agar concentration increases. (b) Temporal velocity ACF broadens as agar concentration increases. (c) Correlation length $L^{\text{corr}}$ and correlation time $T^{\text{corr}}$ as functions of agar concentration $\phi$. (d) Spatial velocity ACF narrows over time. (e) Temporal velocity ACF broadens significantly over time. (f) Correlation length $L^{\text{corr}}$ and correlation time $T^{\text{corr}}$ as functions of time $t$. The dashed reference lines in (a, b, d, e) represent the measurement over the glass substrate. In(a, b), the data are measured $30~\mathrm{s}$ after loading ($t = 0$ min). In (d, e), the data are measured near 0.5 wt\% agar substrates ($\phi = 0.5$ wt\%), and the dashed reference lines show the time-average velocity ACF, with details in [Supplementary Information]. Error bars indicate the standard error.}
    \label{fig:correlation}
\end{figure*}

\textit{Results}.---To distinguish substrate effects on individual swimming from those on collective motion, we compared dilute and dense bacterial suspensions adjacent to glass and agar interfaces. The experimental configurations followed those used in our previous studies \cite{Li2025,bian2026} [see Supplementary Information]. Dilute suspensions of \textit{E.~coli} were confined between a glass coverslip and a lower boundary of either glass or non-nutritive agar, whereas dense suspensions were laterally confined by impermeable sidewalls above either glass or porous agar. Images were acquired at 60~fps, and the in-plane velocity fields were extracted using optical flow method [see Supplementary Information]. The dense suspension had a thickness of approximately $100~\mu\mathrm{m}$, far exceeding the bacterial body length, and therefore permitting three-dimensional motion in the bulk. Our measurements, however, resolved only in-plane velocity fields near the lower interface, at a distance of approximately $10~\mu\mathrm{m}$.

\begin{figure*}[tb]
    \centering
    \includegraphics[width=\linewidth]{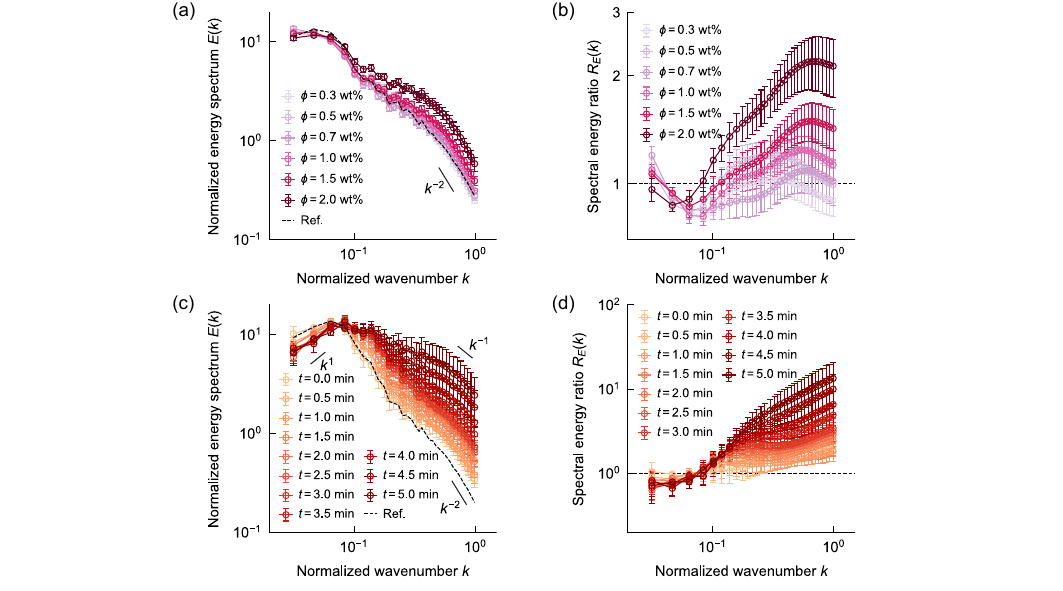}
    \caption{Comparison of energy spectra over glass and agar substrates. The energy spectra $E(k)$ are normalized by kinetic energy and are independent of the activity. The shown wavenumbers $k$ are scaled by $k_b=2\pi/l_b$, where $l_b = 3~\mathrm{\mu m}$ is the typical bacterial body length. (a) Energy spectra over agar substrates in different concentrations for $t = 0$ min. (b) Spectral energy ratio $R_E(k)$ versus agar concentration $\phi$. (c) Energy spectra over time for $\phi = 0.5$ wt\%. (d) Spectral energy ratio $R_E(k)$ versus time $t$. The dashed reference lines in (a, c) represent the measurement over the glass substrate. In (c), the reference energy spectrum is time-averaged. Error bars represent the standard error.}
    \label{fig:energy_spectra}
\end{figure*}

In dilute suspensions, individual bacteria follow circular trajectories, whereas dense suspensions exhibit vortical motion on length scales exceeding the bacterial body length. Both behaviors are consistent with previous studies \cite{Lauga2006, DiLuzio2005, bian2026}. To quantify the dynamics in these two regimes, we define $E_{\mathrm{swim}}=\langle v_{\parallel}\rangle^2/2$ and $E_{\mathrm{turb}}=\langle u^2+v^2\rangle/2$ as measures of individual swimming and collective motion, respectively. Here, $v_{\parallel}=\sqrt{u^2+v^2}$ is the instantaneous in-plane swimming speed, with $u$ and $v$ representing the velocity components along the $x$ and $y$ directions, respectively. Angle brackets denote averages over the corresponding measurements, and both measures exclude the out-of-plane velocity component.

Approximately $30\ \mathrm{s}$ after loading, the normalized kinetic energy for dilute suspensions is similar across different agar concentrations; for concentrated suspensions, however, this value differs by a factor of approximately two [Fig.~\ref{fig:modulation}(a)]. With increasing agar concentrations, the turbulent kinetic energy increases until it reaches a maximum with an agar concentration of about 0.7~wt\%; with further increases in the agar concentration the turbulent kinetic energy gradually decreases. To determine whether this energy variation accompanied a change in characteristic spatial scale, we examined the relationship between energy and enstrophy. We defined the in-plane vorticity as $\omega=\partial v/\partial x-\partial u/\partial y$ and the enstrophy as $\Omega=\langle\omega^{2}\rangle/2$. The ratio of energy and enstrophy $\Lambda^2 = E/\Omega$ defines a gradient-based length scale \cite{Dunkel2013}. We found that the normalized characteristic length $\Lambda$ decreases slightly within a range of approximately 10\% [see Supplementary Information], indicating that the turbulent spatial scale changes much less than the kinetic energy.

To investigate the long-time response beyond the initial concentration-dependent behavior, we followed bacterial turbulence over the first 5~min after the initial measurement. Representative snapshots revealed distinct evolution near glass and 0.5~wt\% agar [Fig.~\ref{fig:modulation}(b)]. For the flow field over the glass, prominent vortical structures persisted throughout the observation period. In contrast, initially pronounced structures weakened markedly within approximately 3~min for the flow field over the agar, as shown by the reduced contrast in the vorticity field. The normalized energy and enstrophy quantified this long-time divergence [Fig.~\ref{fig:modulation}(c)]. Both quantities of the bacterial turbulence over the glass decreased modestly after loading and then remained at comparatively high levels. For the agar, however, both quantities continued to decline and approached zero at the end of the observation period.  

Spatial and temporal velocity autocorrelation functions (ACFs) given by
\begin{equation}
    \begin{split}
        C_r(L) &= \langle \frac{\vb*{v}(\vb*{r}, t) \cdot \vb*{v}(\vb*{r} + L, t)}{|\vb*{v}(\vb*{r}, t)||\vb*{v}(\vb*{r} + L, t)|} \rangle_r, \\
        C_t(T) &= \langle \frac{\vb*{v}(\vb*{r}, t) \cdot \vb*{v}(\vb*{r}, t + T)}{|\vb*{v}(\vb*{r}, t)||\vb*{v}(\vb*{r}, t + T)|} \rangle_t,
    \end{split}
\end{equation}
reveal contrasting responses to different substrates. Further, the correlation length $L^{\text{corr}}$ and correlation time $T^{\text{corr}}$ are defined as $C_r(L^{\text{corr}}) = 1/e$ and $C_t(T^{\text{corr}}) = 1/e$, respectively. As Fig.~\ref{fig:correlation} shows, we characterize the spatial and temporal velocity ACFs at different agar concentrations and over the course of the experiment, using measurements on a glass substrate as a reference. Spatial velocity ACFs exhibit similar decay profiles across different agar concentrations, with a modest acceleration of the decay at higher concentrations [Fig.~\ref{fig:correlation}(a)]. Correspondingly, correlation length \(L^{\mathrm{corr}}\) decreases from approximately 15 $\mu$m at \(\phi=0.3\) wt\% to 13.3 $\mu$m at \(\phi=2.0\) wt\% [Fig.~\ref{fig:correlation}(c)]. In contrast, temporal velocity ACFs become broader with increasing agar concentration [Fig.~\ref{fig:correlation}(b)]. The associated correlation time \(T^{\mathrm{corr}}\) generally increases from approximately 0.3 s to 0.6 s [Fig.~\ref{fig:correlation}(c)], indicating enhanced temporal persistence of the local velocity field. A similar but more pronounced evolution is observed over time. Spatial velocity ACF gradually narrows [Fig.~\ref{fig:correlation}(d)], accompanied by a reduction in \(L^{\mathrm{corr}}\) from approximately 13 $\mu$m to 10 $\mu$m over 5 min [Fig.~\ref{fig:correlation}(f)]. Meanwhile, temporal velocity ACF broadens substantially [Fig.~\ref{fig:correlation}(e)], with \(T^{\mathrm{corr}}\) reaching approximately 1.0 s at later times [Fig.~\ref{fig:correlation}(f)].

Energy spectra provided a further comparison of turbulent flow fields over glass and agar. The kinetic energy spectrum is obtained from the Fourier transformation of the equal-time two-point velocity correlation function under different dimension $d$ \cite{Wensink2012}:
\begin{equation}
    E_d(k) = \frac{k^{d-1}}{C_d}\int d^2 Re^{-i\vb*{k}\cdot\vb*{R}}\langle \vb*{v}(\vb*{r}, t)\cdot \vb*{v}(\vb*{r}+\vb*{R}, t) \rangle,
\end{equation}
where $C_d = 2\pi$ for 2D case and $C_d = 4\pi$ for 3D case. We found that the dimensionless energy spectra measured over different substrates show similar scaling behaviors [Fig.~\ref{fig:energy_spectra}(a)]. Specifically, the energy spectra are featured by two scaling regimes: $E(k)\sim k^{\gamma_1}$ at low-$k$ (large length scales) and $E(k)\sim k^{\gamma_2}$ at high-$k$ (small length scales), where the scaling exponents are approximately $\gamma_1 = 0$ and $\gamma_2 = -2$. Note that the range of values for $\gamma_1$, varying from $-1$ to $1$, may not be strictly accurate due to experimental limitations [see Supplementary Information]. These spectra slightly shift to higher $k$ as $\phi$ increases [Fig.~\ref{fig:energy_spectra}(a)], which agrees with our observation of decreasing $L^{\text{corr}}$ [Fig.~\ref{fig:correlation}(c)]. We also computed the spectral energy ratio $R_E(k) = E(k)/E^{\text{Ref}}(k)$ to compare the energy spectra over agar with over glass. For larger $\phi$, $R_E(k)$ exhibited a higher plateau at high-$k$ [Fig.~\ref{fig:energy_spectra}(b)], indicating that the agar substrate with higher concentration has a stronger modulation in the concentration of kinetic energy at small scales. In contrast, normalized energy spectra measured over the $0.5$ wt\% agar substrate show different scaling behaviors over time, with the high-$k$ scaling exponent $\gamma_2$ changes from $-2$ to $-1$ [Fig.~\ref{fig:energy_spectra}(c)]. Meanwhile, the plateaus at high-$k$ rises over time, with $R_E(k)$ exceeding $10$ at $k \simeq k_b$ at $t = 5$ min [Fig.~\ref{fig:energy_spectra}(d)], suggesting that the agar substrate plays a long-term modulation in transferring kinetic energy from large scales to small scales. The previous measurements on energy spectra are well in line with our results \cite{bian2026, liu_density_2021, Wei2024}.

\begin{figure*}[htb]
    \centering
    \includegraphics[width=\linewidth]{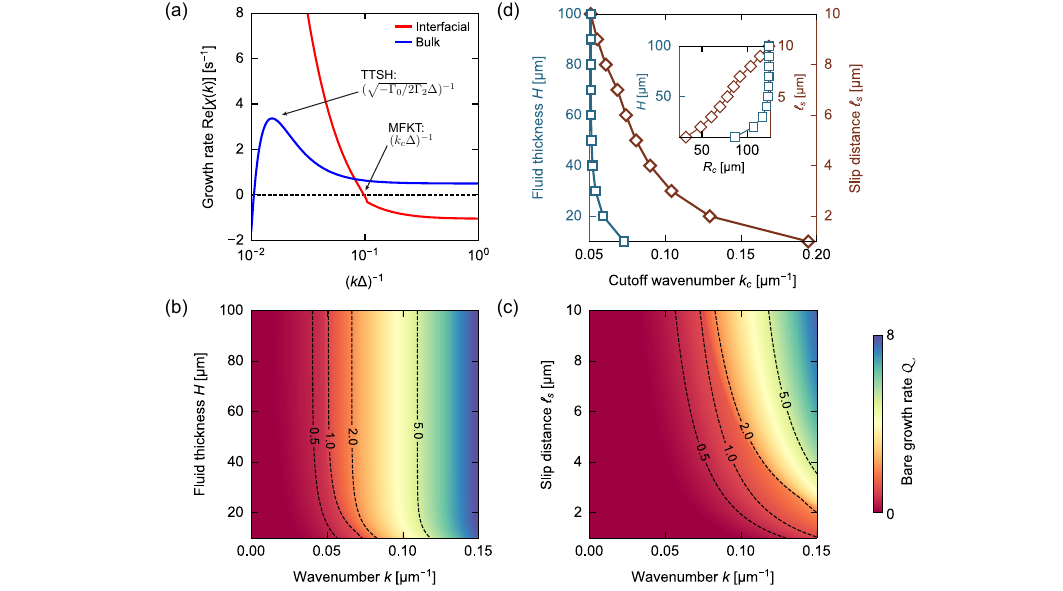}
    \caption{Minimal theoretical framework predicts modulation of the hydrated porous substrate. (a) Growth rate as a function of wavenumber for the active fluid near porous surface and in bulk. Details of parameters are shown in [Supplementary Information]. (b) Contour plot of $\mathcal{Q}$ as a function of the wavenumber $k$ and fluid thickness $H$ with fixed slip distance $\ell_s = 10\ \mathrm{\mu m}$, the color bar indicates the values of $\mathcal{Q}$. Dashed lines represent the onset of instability, $\mathcal{Q} = \lambda$, with contour levels at $0.5$, $1.0$, $2.0$, and $5.0$ $\mathrm{s^{-1}}$. (c) Contour plot of $\mathcal{Q}$ as a function of $k$ and $\ell_s$ with fixed $H = 100\ \mathrm{\mu m}$. (d) Cutoff wavenumber $k_c$ varies as function of $H$ and $\ell_s$. Inset: Cutoff vortex size $R_c$ varies as function of $H$ and $\ell_s$.}
    \label{fig:theory}
\end{figure*}

The shorter spatial correlations and increased relative spectral weight at higher wavenumbers motivate examining how porous boundaries influence collective flow structures. We therefore develop a minimal theoretical framework based on MFKT. The framework incorporates hydrodynamic coupling to the air above the suspension and the porous substrate below. Following previous works \cite{Guillamat2016, MatasNavarro2014, Mathijssen2016, ElKorde2025, PrezEstay2026, Yin2026}, we model bacteria as force dipoles and represent hydrodynamic interactions by an image expansion of the force dipole of strength $\zeta$. The additional terms are introduced due to the tangential Beavers–Joseph partial-slip condition \cite{BeaversJoseph1967} and normal Darcy seepage at the porous surface \cite{Yang2017}, which provides a controlled approximation in the experimentally relevant regime [see Supplementary Information]. 

Linear stability analysis yields the interfacial growth rate as a function of the wavenumber [see Supplementary Information]:
\begin{equation}\label{eq:growth_rate}
    \chi(k) = -\lambda+
  G\left[\frac{2v_s f(kH,k\ell_s)}{B\zeta n}\right]
  \frac{B\zeta nk}{2 f(kH,k\ell_s)},
\end{equation}
where 
$G(x)=6x^2\left[
 4-(1\pm i\sqrt{3})/h(x)
  -(1\mp i\sqrt{3})h(x)
 \right]^{-1}$
and
$h(x)=
 \left[54x^2-1+6\sqrt{3}\,x\sqrt{27x^2-1}\right]^{1/3}$.
Here, $\lambda$ denotes the tumbling rate, $v_s$ represents the bacterial swimming speed, and $B$ is the Bretherton constant accounting for particle shape \cite{Bretherton1962}. The dipole strength is defined as $\zeta = Fl_b/\mu$, where $F$ is the propulsion force and $\mu$ the viscosity of buffer; $n$ represents the interfacial number density. The collective activity establishes a characteristic velocity scale $v_0 = B\zeta n$, with the effects of fluid thickness $H$ and slip distance $\ell_s$ captured by the dimensionless geometric factor $f(x, y) = (\coth x+y) / 4y$. The interplay between active forcing and stochastic reorientation due to tumbling further gives rise to an intrinsic length scale, $\Delta = v_0/\lambda$ \cite{Yin2026}. 

The interfacial dispersion relation shown in Eq. (\ref{eq:growth_rate}) exhibits that the real part of the growth rate, $\operatorname{Re}[\chi(k)]$, increases with increasing wavenumber $k$, producing short-wavelength instability, which is opposite to the bulk fluid with the long-wavelength instability obtained analytically from Toner–Tu–Swift–Hohenberg (TTSH) equation [see Supplementary Information, Fig.~\ref{fig:theory}(a)]. We define the cutoff wavenumber $k_c$ as $\operatorname{Re}[\chi(k_c)] = 0$, and the cutoff vortex size $R_c = 2\pi/k_c$. Although $R_c$ has a much larger value than the characteristic vortex size derived from TTSH equation, it is also a good indicator of the spatial scale of bacterial turbulence \cite{Wei2024, Yin2026}.

To identify the onset of orientational instability, we introduce the bare growth rate $\mathcal{Q}(k, H, \ell_s) = \operatorname{Re}[(v_0k/2f)G(2v_sf/v_0)]$, which characterizes the intrinsic growth of orientational fluctuations in the absence of tumbling ($\lambda=0$), when bacteria follow straight trajectories. The instability threshold is determined by the condition, $\operatorname{Re}[\chi(k)] = 0$, which, according to Eq. (\ref{eq:growth_rate}), reduces to $\mathcal{Q} = \lambda$. This relation highlights the stabilizing role of tumbling, which suppresses the growth of orientational fluctuations and thereby favors the homogeneous isotropic state \cite{SUBRAMANIAN2009}. For fixed $H$ and $\ell_s$, $\mathcal{Q}$ increases rapidly as $k$ increases, suggesting that small vortices appear first [Fig.~\ref{fig:theory}(b, c)].

Theoretical predictions for the cutoff wavenumber $k_c$ and vortex size $R_c$ are given with a typical tumbling rate of \textit{E. coli}, $\lambda = 1.0\ \mathrm{s^{-1}}$ \cite{Lemelle2020}. It shows that decreasing fluid thickness $H$ and slip distance $\ell_s$ causes an increasing $k_c$ (decreasing $R_c$) [Fig.~\ref{fig:theory}(d)]. In our experimental regime, the fluid thickness $H$ continuously decreases over time because of the downward Darcy seepage. On the other hand, the slip distance is given by $\ell_s = \sqrt{K}/\alpha_{BJ}$, where $K$ is the permeability of the substrate and $\alpha_{BJ}$ is a dimensionless coefficient that depends only on the characteristics of the material \cite{Saffman1971}. 
In general, permeability is negatively correlated with concentration, which means that agar with larger $\phi$ has a smaller $\ell_s$ if the change in $\alpha_{BJ}$ is negligible \cite{Gu2003, DelMastro2026}.
Therefore, the qualitative behaviors of $R_c$ are consistent with the experimentally observed decrease of $L^{\text{corr}}$ with $\phi$ and $t$ [Fig.~\ref{fig:correlation}]. This elucidates how a hydrated porous substrate regulates collective interfacial dynamics through the coupling between horizontal and vertical flows, providing a mechanistic explanation for its influence on vortex dynamics.

\textit{Discussion}.---Hydrated porous substrate affects the intensity, persistence, and spatial organization of bacterial turbulence in different ways. The initial nonmonotonic energy response contrasts with the variation in spatial scales, indicating that flow amplitude and characteristic size need not vary together [Fig.~\ref{fig:modulation}(a), Fig.~\ref{fig:correlation}]. Such behavior of kinetic energy is potentially attributed to the competing effects of friction and permeability on collective flows, with the former acting to suppress flow while the latter promotes it. Over five minutes above 0.5~wt\% agar, flow attenuation accompanies shorter spatial correlations and longer temporal correlations [Fig.~\ref{fig:modulation}(c), Fig.~\ref{fig:correlation}]. Normalized spectra also acquire greater relative weight at higher wavenumbers [Fig.~\ref{fig:energy_spectra}]. This shift describes the distribution of remaining motion, rather than demonstrating an increase in absolute small-scale energy or energy transfer between scales.

The extended MFKT connects these spatial changes to hydrodynamic coupling at the porous boundary. Normal seepage and tangential slip modify the stability characteristics of near-interface collective modes relative to those in the bulk. Decreasing fluid-layer thickness and slip length shifts the instability cutoff towards higher wavenumbers, reducing its associated length scale [Fig.~\ref{fig:theory}]. These trends provide a qualitative interpretation of the smaller flow structures observed over time and at higher agar concentrations. However, the cutoff length is distinct from the measured correlation length, and linear stability alone does not predict the nonmonotonic energy response or subsequent attenuation. The concentration also changes several substrate properties together, preventing their individual contributions from being isolated. Measurements of fluid thickness, permeability, and interfacial slip would help establish their correspondence with the model parameters.

These findings may inform studies of swarming, where collective motion depends on surface hydration \cite{kearns_field_2010,Jose2025LiquidDepletion}. Although the thicker suspensions studied here differ from swarming layers, both systems motivate understanding how hydrated porous substrates influence collective motion. Our experiments show that hydrated porous substrates modify the spatiotemporal dynamics of near-interface bacterial turbulence, with distinct concentration-dependent and temporal responses. Minimal theoretical framework identifies normal seepage and tangential slip as hydrodynamic contributions that can modify collective-mode stability and associated spatial scales. Together, these results establish hydrated porous substrates as an important boundary condition for understanding and regulating bacterial collective flows.

\textit{Acknowledgments}.---We thank Y. Huang for the optical flow codes. We are also grateful to K. Chang for fruitful discussions. The authors declare no financial support was received for this work.


\bibliography{references}

@article{Wensink2012,
  title = {Meso-scale turbulence in living fluids},
  volume = {109},
  ISSN = {1091-6490},
  url = {http://dx.doi.org/10.1073/pnas.1202032109},
  DOI = {10.1073/pnas.1202032109},
  number = {36},
  journal = {Proceedings of the National Academy of Sciences},
  publisher = {National Academy of Sciences},
  author = {Wensink,  Henricus H. and Dunkel,  J\"{o}rn and Heidenreich,  Sebastian and Drescher,  Knut and Goldstein,  Raymond E. and L\"{o}wen,  Hartmut and Yeomans,  Julia M.},
  year = {2012},
  month = Aug,
  pages = {14308–14313}
}

@article{Sokolov2007,
  title = {Concentration Dependence of the Collective Dynamics of Swimming Bacteria},
  volume = {98},
  ISSN = {1079-7114},
  url = {http://dx.doi.org/10.1103/PhysRevLett.98.158102},
  DOI = {10.1103/physrevlett.98.158102},
  number = {15},
  journal = {Physical Review Letters},
  publisher = {American Physical Society (APS)},
  author = {Sokolov,  Andrey and Aranson,  Igor S. and Kessler,  John O. and Goldstein,  Raymond E.},
  year = {2007},
  month = Apr,
  pages = {158102},
  numpages = {4}
}

@article{annurev-ActiveTurbulence,
   author = "Alert, Ricard and Casademunt, Jaume and Joanny, Jean-François",
   title = "Active Turbulence", 
   journal= "Annual Review of Condensed Matter Physics",
   year = "2022",
   volume = "13",
   number = "Volume 13, 2022",
   pages = "143-170",
   doi = {10.1146/annurev-conmatphys-082321-035957},
   url = "https://www.annualreviews.org/content/journals/10.1146/annurev-conmatphys-082321-035957",
   publisher = "Annual Reviews",
   issn = "1947-5462",
   type = "Journal Article",
}

@article{zhang_swarming_2009,
	title = {Swarming dynamics in bacterial colonies},
	volume = {87},
	issn = {0295-5075, 1286-4854},
	url = {https://iopscience.iop.org/article/10.1209/0295-5075/87/48011},
	doi = {10.1209/0295-5075/87/48011},
	number = {4},
	urldate = {2024-06-18},
	journal = {EPL (Europhysics Letters)},
	author = {Zhang, H. P. and Be'er, Avraham and Smith, Rachel S. and Florin, E.-L. and Swinney, Harry L.},
	month = aug,
	year = {2009},
	pages = {48011},
}

@article{kearns_field_2010,
	title = {A field guide to bacterial swarming motility},
	volume = {8},
	copyright = {http://www.springer.com/tdm},
	issn = {1740-1526, 1740-1534},
	number = {9},
	urldate = {2024-06-13},
	journal = {Nature Reviews Microbiology},
	author = {Kearns, Daniel B.},
	doi = {10.1038/nrmicro2405},
	month = sep,
	year = {2010},
	pages = {634--644},
}

@article{beer_statistical_2019,
	title = {A statistical physics view of swarming bacteria},
	volume = {7},
	issn = {2051-3933},
	url = {https://movementecologyjournal.biomedcentral.com/articles/10.1186/s40462-019-0153-9},
	doi = {10.1186/s40462-019-0153-9},
	number = {1},
	urldate = {2024-05-16},
	journal = {Movement Ecology},
	author = {Be’er, Avraham and Ariel, Gil},
	month = dec,
	year = {2019},
	pages = {9},
}

@article{peng_imaging_2021,
	title = {Imaging the emergence of bacterial turbulence: {Phase} diagram and transition kinetics},
	volume = {7},
	copyright = {https://creativecommons.org/licenses/by-nc/4.0/},
	issn = {2375-2548},
	shorttitle = {Imaging the emergence of bacterial turbulence},
	url = {https://www.science.org/doi/10.1126/sciadv.abd1240},
	doi = {10.1126/sciadv.abd1240},
	number = {17},
	urldate = {2025-04-08},
	journal = {Science Advances},
	author = {Peng, Yi and Liu, Zhengyang and Cheng, Xiang},
	month = apr,
	year = {2021},
	pages = {eabd1240},
}

@article{liu_density_2021,
	title = {Density fluctuations and energy spectra of {3D} bacterial suspensions},
	volume = {17},
	issn = {1744-683X, 1744-6848},
	url = {https://xlink.rsc.org/?DOI=D1SM01183A},
	doi = {10.1039/D1SM01183A},
	number = {48},
	urldate = {2025-04-27},
	journal = {Soft Matter},
	author = {Liu, Zhengyang and Zeng, Wei and Ma, Xiaolei and Cheng, Xiang},
	year = {2021},
	pages = {10806--10817},
}

@article{bian2026,
  title = {Active turbulence in shear-thinning fluid},
  author = {Bian, Hongyi and Li, Chunhe and Lin, Zixiang and Zhu, Jin and Chen, Weijie and Li, Gaojin and Huang, Yongxiang and Qu, Zijie},
  journal = {Phys. Rev. Res.},
  volume = {8},
  issue = {1},
  pages = {013318},
  numpages = {9},
  year = {2026},
  month = {Mar},
  publisher = {American Physical Society},
  doi = {10.1103/z22f-3v22},
  url = {https://link.aps.org/doi/10.1103/z22f-3v22}
}

@article{Li2025,
  title = {Hydrodynamic interaction leads to the accumulation of Chlamydomonas reinhardtii near a solid-liquid interface},
  author = {Li, Chunhe and Bian, Hongyi and Qiao, Yateng and Zhu, Jin and Qu, Zijie},
  journal = {Phys. Rev. E},
  volume = {111},
  issue = {5},
  pages = {L052401},
  numpages = {6},
  year = {2025},
  month = {May},
  publisher = {American Physical Society},
  doi = {10.1103/PhysRevE.111.L052401},
  url = {https://link.aps.org/doi/10.1103/PhysRevE.111.L052401}
}

@article{Dunkel2013,
  title = {Fluid Dynamics of Bacterial Turbulence},
  volume = {110},
  ISSN = {1079-7114},
  url = {http://dx.doi.org/10.1103/PhysRevLett.110.228102},
  DOI = {10.1103/physrevlett.110.228102},
  number = {22},
  journal = {Physical Review Letters},
  publisher = {American Physical Society (APS)},
  author = {Dunkel,  J\"{o}rn and Heidenreich,  Sebastian and Drescher,  Knut and Wensink,  Henricus H. and B\"{a}r,  Markus and Goldstein,  Raymond E.},
  year = {2013},
  pages = {228102},
  month = May 
}

@article{spera_lowpass_2026,
  author  = {Spera, Gianmarco and Yeomans, Julia M. and Thampi, Sumesh P.},
  title   = {Low-pass filtering of active turbulent flows to liquid substrates},
  journal = {Newton},
  year    = {2026},
  volume  = {2},
  number  = {8},
  pages   = {100524},
  doi     = {10.1016/j.newton.2026.100524},
  url     = {https://doi.org/10.1016/j.newton.2026.100524}
}

@article{mecke_chiral_2024,
  author  = {Mecke, Joscha and Gao, Yongxiang and Gompper, Gerhard and Ripoll, Marisol},
  title   = {Chiral active systems near a substrate: Emergent damping length controlled by fluid friction},
  journal = {Communications Physics},
  year    = {2024},
  volume  = {7},
  number  = {1},
  pages   = {332},
  doi     = {10.1038/s42005-024-01817-0},
  url     = {https://doi.org/10.1038/s42005-024-01817-0}
}

@article{Jose2025LiquidDepletion,
  author  = {Jose, Ajesh and P{\'e}rez-Estay, Benjam{\'i}n and Bendori, Shira Omer and Eldar, Avigdor and Kearns, Daniel B. and Ariel, Gil and Be'er, Avraham},
  title   = {Immobility of isolated swarmer cells due to local liquid depletion},
  journal = {Communications Physics},
  year    = {2025},
  volume  = {8},
  number  = {1},
  pages   = {88},
  doi     = {10.1038/s42005-025-01996-4}
}

@article{thijssen_anisotropic_2020,
  author  = {Thijssen, Kristian and Metselaar, Luuk and Yeomans, Julia M. and Doostmohammadi, Amin},
  title   = {Active nematics with anisotropic friction: the decisive role of the flow aligning parameter},
  journal = {Soft Matter},
  year    = {2020},
  volume  = {16},
  number  = {8},
  pages   = {2065--2074},
  doi     = {10.1039/C9SM01963D},
  url     = {https://doi.org/10.1039/C9SM01963D}
}

@article{pokhrel_friction_2025,
  author  = {Pokhrel, Aawaz R. and Copeland, Raymond and Hejri, Maryam and Belpaire, Tom E. R. and Steinbach, Gabi and Ng, Siu Lung and Hammer, Brian K. and Yunker, Peter J.},
  title   = {Cell-substrate friction controls biofilm development},
  journal = {bioRxiv},
  year    = {2025},
  doi     = {10.1101/2025.07.11.664457},
  url     = {https://www.biorxiv.org/content/10.1101/2025.07.11.664457v1},
  note    = {Preprint}
}

@article{beer_colony_2009,
  author  = {Be'er, Avraham and Smith, Rachel S. and Zhang, H. P. and Florin, E.-L. and Payne, Shelley M. and Swinney, Harry L.},
  title   = {{Paenibacillus dendritiformis} bacterial colony growth depends on surfactant but not on bacterial motion},
  journal = {Journal of Bacteriology},
  year    = {2009},
  volume  = {191},
  number  = {18},
  pages   = {5758--5764},
  doi     = {10.1128/JB.00660-09},
  url     = {https://doi.org/10.1128/JB.00660-09}
}

@article{Dunkel2013M,
  title = {Minimal continuum theories of structure formation in dense active fluids},
  volume = {15},
  ISSN = {1367-2630},
  url = {http://dx.doi.org/10.1088/1367-2630/15/4/045016},
  DOI = {10.1088/1367-2630/15/4/045016},
  number = {4},
  journal = {New Journal of Physics},
  publisher = {IOP Publishing},
  author = {Dunkel,  J\"{o}rn and Heidenreich,  Sebastian and B\"{a}r,  Markus and Goldstein,  Raymond E},
  year = {2013},
  month = Apr,
  pages = {045016}
}

@article{BeaversJoseph1967,
  author  = {Beavers, Gordon S. and Joseph, Daniel D.},
  title   = {Boundary conditions at a naturally permeable wall},
  journal = {Journal of Fluid Mechanics},
  year    = {1967},
  volume  = {30},
  number  = {1},
  pages   = {197--207},
  doi     = {10.1017/S0022112067001375}
}

@article{Saffman1971,
  author  = {Saffman, P. G.},
  title   = {On the boundary condition at the surface of a porous medium},
  journal = {Studies in Applied Mathematics},
  year    = {1971},
  volume  = {50},
  number  = {2},
  pages   = {93--101},
  doi     = {10.1002/sapm197150293}
}

@article{Lauga2006,
  title = {Swimming in Circles: Motion of Bacteria near Solid Boundaries},
  volume = {90},
  ISSN = {0006-3495},
  url = {http://dx.doi.org/10.1529/biophysj.105.069401},
  DOI = {10.1529/biophysj.105.069401},
  number = {2},
  journal = {Biophysical Journal},
  publisher = {Elsevier BV},
  author = {Lauga,  Eric and DiLuzio,  Willow R. and Whitesides,  George M. and Stone,  Howard A.},
  year = {2006},
  month = Jan,
  pages = {400–412}
}

@article{Wei2024,
  title = {Scaling Transition of Active Turbulence from Two to Three Dimensions},
  volume = {11},
  ISSN = {2198-3844},
  url = {http://dx.doi.org/10.1002/advs.202402643},
  DOI = {10.1002/advs.202402643},
  number = {38},
  journal = {Advanced Science},
  publisher = {Wiley},
  author = {Wei,  Da and Yang,  Yaochen and Wei,  Xuefeng and Golestanian,  Ramin and Li,  Ming and Meng,  Fanlong and Peng,  Yi},
  year = {2024},
  pages = {2402643},
  month = Aug 
}

@article{Marchetti2013,
  title = {Hydrodynamics of soft active matter},
  volume = {85},
  ISSN = {1539-0756},
  url = {http://dx.doi.org/10.1103/RevModPhys.85.1143},
  DOI = {10.1103/revmodphys.85.1143},
  number = {3},
  journal = {Reviews of Modern Physics},
  publisher = {American Physical Society (APS)},
  author = {Marchetti,  M. C. and Joanny,  J. F. and Ramaswamy,  S. and Liverpool,  T. B. and Prost,  J. and Rao,  Madan and Simha,  R. Aditi},
  year = {2013},
  month = July,
  pages = {1143–1189}
}

@article{Guillamat2016,
  title = {Probing the shear viscosity of an active nematic film},
  volume = {94},
  ISSN = {2470-0053},
  url = {http://dx.doi.org/10.1103/PhysRevE.94.060602},
  DOI = {10.1103/physreve.94.060602},
  number = {6},
  journal = {Physical Review E},
  publisher = {American Physical Society (APS)},
  author = {Guillamat,  Pau and Ignés-Mullol,  Jordi and Shankar,  Suraj and Marchetti,  M. Cristina and Sagués,  Francesc},
  year = {2016},
  pages = {060602},
  month = Dec 
}

@article{MartnezPrat2021,
  title = {Scaling Regimes of Active Turbulence with External Dissipation},
  volume = {11},
  ISSN = {2160-3308},
  url = {http://dx.doi.org/10.1103/PhysRevX.11.031065},
  DOI = {10.1103/physrevx.11.031065},
  number = {3},
  journal = {Physical Review X},
  publisher = {American Physical Society (APS)},
  author = {Martínez-Prat,  Berta and Alert,  Ricard and Meng,  Fanlong and Ignés-Mullol,  Jordi and Joanny,  Jean-Fran\c{c}ois and Casademunt,  Jaume and Golestanian,  Ramin and Sagués,  Francesc},
  year = {2021},
  pages = {031065},
  month = Sept 
}

@article{Yin2026,
  title = {Bacterial Turbulence at Compressible Fluid Interfaces},
  volume = {136},
  ISSN = {1079-7114},
  url = {http://dx.doi.org/10.1103/ww6w-f94m},
  DOI = {10.1103/ww6w-f94m},
  number = {13},
  journal = {Physical Review Letters},
  publisher = {American Physical Society (APS)},
  author = {Yin,  Yuanfeng and Zhang,  Bokai and Zhang,  H. P. and Guo,  Shuo},
  year = {2026},
  pages = {138301},
  month = Mar 
}

@article{MatasNavarro2014,
  title = {Hydrodynamic suppression of phase separation in active suspensions},
  volume = {90},
  ISSN = {1550-2376},
  url = {http://dx.doi.org/10.1103/PhysRevE.90.032304},
  DOI = {10.1103/physreve.90.032304},
  number = {3},
  journal = {Physical Review E},
  publisher = {American Physical Society (APS)},
  author = {Matas-Navarro,  Ricard and Golestanian,  Ramin and Liverpool,  Tanniemola B. and Fielding,  Suzanne M.},
  year = {2014},
  pages = {032304},
  month = Sept 
}

@article{Mathijssen2016,
  title = {Hydrodynamics of micro-swimmers in films},
  volume = {806},
  ISSN = {1469-7645},
  url = {http://dx.doi.org/10.1017/jfm.2016.479},
  DOI = {10.1017/jfm.2016.479},
  journal = {Journal of Fluid Mechanics},
  publisher = {Cambridge University Press (CUP)},
  author = {Mathijssen,  A. J. T. M. and Doostmohammadi,  A. and Yeomans,  J. M. and Shendruk,  T. N.},
  year = {2016},
  month = Sept,
  pages = {35–70}
}

@article{kultty2024,
  title = {Hydrodynamic instabilities in a two-dimensional sheet of microswimmers embedded in a three-dimensional fluid},
  volume = {980},
  ISSN = {1469-7645},
  url = {http://dx.doi.org/10.1017/jfm.2023.985},
  DOI = {10.1017/jfm.2023.985},
  journal = {Journal of Fluid Mechanics},
  publisher = {Cambridge University Press (CUP)},
  author = {Škultéty,  Viktor and Bárdfalvy,  Dóra and Stenhammar,  Joakim and Nardini,  Cesare and Morozov,  Alexander},
  year = {2024},
  month = Feb,
  pages={A28}
}

@article{Bretherton1962,
  title = {The motion of rigid particles in a shear flow at low Reynolds number},
  volume = {14},
  ISSN = {1469-7645},
  url = {http://dx.doi.org/10.1017/S002211206200124X},
  DOI = {10.1017/s002211206200124x},
  number = {2},
  journal = {Journal of Fluid Mechanics},
  publisher = {Cambridge University Press (CUP)},
  author = {Bretherton,  F. P.},
  year = {1962},
  month = Oct,
  pages = {284–304}
}

@article{Yang2017,
  title = {Influence of Physical Effects on the Swarming Motility of Pseudomonas aeruginosa},
  volume = {112},
  ISSN = {0006-3495},
  url = {http://dx.doi.org/10.1016/j.bpj.2017.02.019},
  DOI = {10.1016/j.bpj.2017.02.019},
  number = {7},
  journal = {Biophysical Journal},
  publisher = {Elsevier BV},
  author = {Yang,  Alexander and Tang,  Wai Shing and Si,  Tieyan and Tang,  Jay X.},
  year = {2017},
  month = Apr,
  pages = {1462–1471}
}

@article{SUBRAMANIAN2009,
  title = {Critical bacterial concentration for the onset of collective swimming},
  volume = {632},
  ISSN = {1469-7645},
  url = {http://dx.doi.org/10.1017/S002211200900706X},
  DOI = {10.1017/s002211200900706x},
  journal = {Journal of Fluid Mechanics},
  publisher = {Cambridge University Press (CUP)},
  author = {SUBRAMANIAN,  GANESH and KOCH,  DONALD L.},
  year = {2009},
  month = July,
  pages = {359–400}
}

@article{Lemelle2020,
  title = {Tumble Kinematics of Escherichia coli near a Solid Surface},
  volume = {118},
  ISSN = {0006-3495},
  url = {http://dx.doi.org/10.1016/j.bpj.2020.03.024},
  DOI = {10.1016/j.bpj.2020.03.024},
  number = {10},
  journal = {Biophysical Journal},
  publisher = {Elsevier BV},
  author = {Lemelle,  Laurence and Cajgfinger,  Thomas and Nguyen,  Cao Cuong and Dominjon,  Agnès and Place,  Christophe and Chatre,  Elodie and Barbier,  Rémi and Palierne,  Jean-Fran\c{c}ois and Vaillant,  Cédric},
  year = {2020},
  month = May,
  pages = {2400–2410}
}

@article{nishiguchi_vortex_2025,
  title   = {Vortex reversal is a precursor of confined bacterial turbulence},
  author  = {Nishiguchi, Daiki and Shiratani, Sora and Takeuchi, Kazumasa A. and Aranson, Igor S.},
  journal = {Proceedings of the National Academy of Sciences},
  year    = {2025},
  volume  = {122},
  number  = {11},
  pages   = {e2414446122},
  doi     = {10.1073/pnas.2414446122}
}

@article{puggioni_giant_2022,
  title   = {Giant vortex dynamics in confined bacterial turbulence},
  author  = {Puggioni, L. and Boffetta, G. and Musacchio, S.},
  journal = {Physical Review E},
  year    = {2022},
  volume  = {106},
  number  = {5},
  pages   = {055103},
  doi     = {10.1103/PhysRevE.106.055103}
}

@article{Sokolov2012,
  title = {Physical Properties of Collective Motion in Suspensions of Bacteria},
  volume = {109},
  ISSN = {1079-7114},
  url = {http://dx.doi.org/10.1103/PhysRevLett.109.248109},
  DOI = {10.1103/physrevlett.109.248109},
  number = {24},
  journal = {Physical Review Letters},
  publisher = {American Physical Society (APS)},
  author = {Sokolov,  Andrey and Aranson,  Igor S.},
  year = {2012},
  month = Dec,
  pages = {248109},
  numpages = {5}
}

@article{Brdfalvy2024,
  title = {Collective motion in a sheet of microswimmers},
  volume = {7},
  ISSN = {2399-3650},
  url = {http://dx.doi.org/10.1038/s42005-024-01587-9},
  DOI = {10.1038/s42005-024-01587-9},
  number = {1},
  journal = {Communications Physics},
  publisher = {Springer Science and Business Media LLC},
  author = {Bárdfalvy,  Dóra and Škultéty,  Viktor and Nardini,  Cesare and Morozov,  Alexander and Stenhammar,  Joakim},
  year = {2024},
  month = Mar,
  pages = {93}
}

@article{Wioland2013,
  title = {Confinement Stabilizes a Bacterial Suspension into a Spiral Vortex},
  volume = {110},
  ISSN = {1079-7114},
  url = {http://dx.doi.org/10.1103/PhysRevLett.110.268102},
  DOI = {10.1103/physrevlett.110.268102},
  number = {26},
  journal = {Physical Review Letters},
  publisher = {American Physical Society (APS)},
  author = {Wioland,  Hugo and Woodhouse,  Francis G. and Dunkel,  J\"{o}rn and Kessler,  John O. and Goldstein,  Raymond E.},
  year = {2013},
  month = June,
  pages = {268102},
  numpages = {5}
}

@article{Wioland2016,
  title = {Directed collective motion of bacteria under channel confinement},
  volume = {18},
  ISSN = {1367-2630},
  url = {http://dx.doi.org/10.1088/1367-2630/18/7/075002},
  DOI = {10.1088/1367-2630/18/7/075002},
  number = {7},
  journal = {New Journal of Physics},
  publisher = {IOP Publishing},
  author = {Wioland,  H and Lushi,  E and Goldstein,  R E},
  year = {2016},
  month = July,
  pages = {075002}
}

@article{Beppu2021,
  title = {Edge current and pairing order transition in chiral bacterial vortices},
  volume = {118},
  ISSN = {1091-6490},
  url = {http://dx.doi.org/10.1073/pnas.2107461118},
  DOI = {10.1073/pnas.2107461118},
  number = {39},
  journal = {Proceedings of the National Academy of Sciences},
  publisher = {National Academy of Sciences},
  author = {Beppu,  Kazusa and Izri,  Ziane and Sato,  Tasuku and Yamanishi,  Yoko and Sumino,  Yutaka and Maeda,  Yusuke T.},
  year = {2021},
  month = Sept,
  pages = {e2107461118}
}

@article{Henshaw2023,
  title = {Dynamic mode structure of active turbulence},
  volume = {8},
  ISSN = {2469-990X},
  url = {http://dx.doi.org/10.1103/PhysRevFluids.8.023101},
  DOI = {10.1103/physrevfluids.8.023101},
  number = {2},
  journal = {Physical Review Fluids},
  publisher = {American Physical Society (APS)},
  author = {Henshaw,  Richard J. and Martin,  Olivia G. and Guasto,  Jeffrey S.},
  year = {2023},
  month = Feb,
  pages = {023101},
  numpages = {15}
}

@article{Nishiguchi2018,
  title = {Engineering bacterial vortex lattice via direct laser lithography},
  volume = {9},
  ISSN = {2041-1723},
  url = {http://dx.doi.org/10.1038/s41467-018-06842-6},
  DOI = {10.1038/s41467-018-06842-6},
  number = {1},
  journal = {Nature Communications},
  publisher = {Springer Science and Business Media LLC},
  author = {Nishiguchi,  Daiki and Aranson,  Igor S and Snezhko,  Alexey and Sokolov,  Andrey},
  year = {2018},
  month = Oct,
  pages = {4486}
}

@article{Reinken2020,
  title = {Organizing bacterial vortex lattices by periodic obstacle arrays},
  volume = {3},
  ISSN = {2399-3650},
  url = {http://dx.doi.org/10.1038/s42005-020-0337-z},
  DOI = {10.1038/s42005-020-0337-z},
  number = {1},
  journal = {Communications Physics},
  publisher = {Springer Science and Business Media LLC},
  author = {Reinken,  Henning and Nishiguchi,  Daiki and Heidenreich,  Sebastian and Sokolov,  Andrey and B\"{a}r,  Markus and Klapp,  Sabine H. L. and Aranson,  Igor S.},
  year = {2020},
  month = May,
  pages = {76}
}

@article{Boffetta2005,
  title = {Effects of friction on 2D turbulence: An experimental study of the direct cascade},
  volume = {71},
  ISSN = {1286-4854},
  url = {http://dx.doi.org/10.1209/epl/i2005-10111-6},
  DOI = {10.1209/epl/i2005-10111-6},
  number = {4},
  journal = {Europhysics Letters (EPL)},
  publisher = {IOP Publishing},
  author = {Boffetta,  G and Cenedese,  A and Espa,  S and Musacchio,  S},
  year = {2005},
  month = Aug,
  pages = {590–596}
}

@article{kultty2020,
  title = {Swimming Suppresses Correlations in Dilute Suspensions of Pusher Microorganisms},
  volume = {10},
  ISSN = {2160-3308},
  url = {http://dx.doi.org/10.1103/PhysRevX.10.031059},
  DOI = {10.1103/physrevx.10.031059},
  number = {3},
  journal = {Physical Review X},
  publisher = {American Physical Society (APS)},
  author = {Škultéty,  Viktor and Nardini,  Cesare and Stenhammar,  Joakim and Marenduzzo,  Davide and Morozov,  Alexander},
  year = {2020},
  month = Sept,
  pages = {031059},
  numpages = {24}
}

@article{PrezEstay2026,
  title = {Collective motion in bacterial suspensions is scale-free},
  volume = {123},
  ISSN = {1091-6490},
  url = {http://dx.doi.org/10.1073/pnas.2600266123},
  DOI = {10.1073/pnas.2600266123},
  number = {21},
  journal = {Proceedings of the National Academy of Sciences},
  publisher = {National Academy of Sciences},
  author = {Pérez-Estay,  Benjamín and Martinez,  Vincent and Douarche,  Carine and Schwarz-Linek,  Jana and Arlt,  Jochen and Delville,  Pierre-Henri and McConnell,  Gail and Poon,  Wilson C. K. and Lindner,  Anke and Clement,  Eric},
  year = {2026},
  month = May,
  pages = {e2600266123}
}

@article{ElKorde2025,
  title = {Pseudogiant Number Fluctuations and Nematic Order in Microswimmer Suspensions},
  volume = {135},
  ISSN = {1079-7114},
  url = {http://dx.doi.org/10.1103/31q1-hj3n},
  DOI = {10.1103/31q1-hj3n},
  number = {9},
  journal = {Physical Review Letters},
  publisher = {American Physical Society (APS)},
  author = {El Korde,  Ismail and Bárdfalvy,  Dóra and Lewis,  Jason M. and Morozov,  Alexander and Nardini,  Cesare and Stenhammar,  Joakim},
  year = {2025},
  month = Aug,
  pages = {098302},
  numpages = {8}
}

@article{DiLuzio2005,
  title = {Escherichia coli swim on the right-hand side},
  volume = {435},
  ISSN = {1476-4687},
  url = {http://dx.doi.org/10.1038/nature03660},
  DOI = {10.1038/nature03660},
  number = {7046},
  journal = {Nature},
  publisher = {Springer Science and Business Media LLC},
  author = {DiLuzio,  Willow R. and Turner,  Linda and Mayer,  Michael and Garstecki,  Piotr and Weibel,  Douglas B. and Berg,  Howard C. and Whitesides,  George M.},
  year = {2005},
  month = June,
  pages = {1271–1274}
}

@article{Li2024,
  title = {Robust Edge Flows in Swarming Bacterial Colonies},
  volume = {14},
  ISSN = {2160-3308},
  url = {http://dx.doi.org/10.1103/PhysRevX.14.041006},
  DOI = {10.1103/physrevx.14.041006},
  number = {4},
  journal = {Physical Review X},
  publisher = {American Physical Society (APS)},
  author = {Li,  He and Chaté,  Hugues and Sano,  Masaki and Shi,  Xia-qing and Zhang,  H. P.},
  year = {2024},
  month = Oct,
  pages = {041006},
  numpages = {15}
}

@article{DelMastro2026,
  title = {Scale-Dependent Slip at Porous Interfaces: Micro-PIV Measurements of Beavers–Joseph and Saffman Coefficients},
  volume = {3},
  ISSN = {3105-6369},
  url = {http://dx.doi.org/10.69631/gexq8215},
  DOI = {10.69631/gexq8215},
  number = {3},
  journal = {InterPore Journal},
  publisher = {The International Society for Porous Media (InterPore)},
  author = {Del Mastro,  Mario and Terzis,  Alexandros},
  year = {2026},
  month = Sept,
  pages = {IPJ020926–3}
}

@article{Gu2003,
  title = {New insight into deformation-dependent hydraulic permeability of gels and cartilage,  and dynamic behavior of agarose gels in confined compression},
  volume = {36},
  ISSN = {0021-9290},
  url = {http://dx.doi.org/10.1016/S0021-9290(02)00437-2},
  DOI = {10.1016/s0021-9290(02)00437-2},
  number = {4},
  journal = {Journal of Biomechanics},
  publisher = {Elsevier BV},
  author = {Gu,  W.Y. and Yao,  H. and Huang,  C.Y. and Cheung,  H.S.},
  year = {2003},
  month = Apr,
  pages = {593–598}
}


\end{document}